\documentclass[aps,pre,floats,superscriptaddress,nofootinbib,floatfix,twocolumn,showkeys]{revtex4-1}
\usepackage[T1]{fontenc}
\usepackage{mathpazo}
\usepackage{amssymb,amsmath,amsfonts,amsthm}
\usepackage{graphicx}
\usepackage{dcolumn}
\usepackage{bm}
\usepackage{comment}
\usepackage{csquotes}
\usepackage{soul}
\usepackage{hyperref}
\usepackage{makecell}
\usepackage{changepage}
\usepackage{multirow}
\usepackage[table]{xcolor}
\usepackage[mathscr]{euscript}

\begin{document}

\title{Agent-Based simulation reveals localized isolation key to saving lives and resources}
\author{Mintu Karmakar}
\email{mcsmk2057@iacs.res.in}
\affiliation{School of Mathematical \& Computational Sciences,\\ Indian Association for the Cultivation of Science, Kolkata -- 700032, India.}

\begin{abstract} 
In the realm of pandemic dynamics, understanding the intricate interplay between disease transmission, interventions, and immunity is pivotal for effective control strategies. Through a rigorous agent-based computer simulation, we embarked on a comprehensive exploration, traversing unmitigated spread, lockdown scenarios, and the transformative potential of vaccination. We unveil that while quarantine unquestionably delays the pandemic peak, it does not act as an impenetrable barrier to halt the progression of infectious diseases. Vaccination factor revealed a potent weapon against outbreaks – higher vaccination percentage not only delayed infection peaks but also substantially curtailed their impact. Our investigation into bond dilution below the percolation threshold presents an additional dimension to pandemic control. We observed that localized isolation through bond dilution offers a more resource-efficient targeted control strategy than blanket lockdowns or large-scale vaccination campaigns. 
\end{abstract}

\maketitle

\section{Introduction}\label{intro}
In the realm of infectious disease control~\cite{molyneux2004, o2002shining, qualls2017community, morse1995factors}, the ongoing effort to find effective methods for curbing the rapid transmission of pandemics remains an enduring challenge. Among the arsenal of measures deployed, quarantine has gained prominence as a key approach, aiming to control the spread of infection by isolating those who are infected from those who are susceptible~\cite{bogoch2015assessment, ferguson2006strategies, nussbaumer2020quarantine}. Yet, as the world confronts repeated instances of pandemic outbreaks~\cite{morens2013emerging, morse1995factors, taubenberger20061918, gostin2020governmental, khan2009spread}, a paradoxical narrative has begun to emerge – one that suggests that while quarantine may act as a crucial temporal barrier, it might not offer a complete solution to halt the unyielding progression of infectious diseases~\cite{patel2020,lakhal2023,wells2021optimal}.

Throughout the history of pandemic control~\cite{piret2021, VILCHES2021106564, Eugenia2013}, quarantine has been recognized as a significant method~\cite{bogoch2015assessment, ferguson2006strategies, nussbaumer2020quarantine}, with its roots dating back to medieval periods when isolated communities were cordoned off to counteract the impacts of the Black Death~\cite{na2016black}. In the present era, this concept has developed into targeted isolation, limitations on travel, and localized shutdowns, all designed to interrupt the progression of transmission. The fundamental principle underpinning quarantine is its ability to impede the rapid proliferation of infections~\cite{nussbaumer2020quarantine, ferguson2006strategies, Chatterjee, Chatterjee1}, affording healthcare systems a period of respite to mobilize resources and formulate more effective countermeasures~\cite{okumura2019, tanaka2019, rocklov2020}.

However, the nature of pandemics~\cite{david2009}, with their ability to cross geographical boundaries and advance across populations, challenges the idea of a complete halt or lockdown through quarantine alone. In the context of the COVID-19 pandemic, several recent studies explored the effectiveness of the lockdown in controlling the disease spread~\cite{vinceti2020lockdown, oraby2021modeling, grimm2021extensions, pei2020differential}. While a lockdown serves as an efficient initial approach to managing the pandemic, it inevitably leads to significant societal and economic repercussions. The critical task for governments across the globe has been to devise epidemic control strategies that ensure public health safety without compromising socio-economic stability. The COVID-19 pandemic triggered the sharpest downturn in the world economy since the Great Depression, with global GDP declining 3.0 percent in 2020 compared to a rise of 2.8 percent in 2019 (IMF 2022)~\cite{imf2022world}. A recent study found that the intensity of lockdown measures—rather than the death toll—plays a more crucial role in affecting GDP growth, particularly in economically underprivileged nations~\cite{gagnon2023impact}. In light of these findings, a key debate revolves around finding a balanced approach to lifting lockdown measures. The ideal strategy would ensure continued health safety without overloading healthcare systems, thereby balancing public health and economic vitality.

To shed light on the subtle interplay between quarantine and the course of an outbreak, this study employs a numerical simulation approach based on the Susceptible-Infectious-Recovered-Deceased (SIRD) model~\cite{KM1927, KM1932}. While the SIR model is simpler, the SIRD model provides a more nuanced and comprehensive framework for analyzing the outcomes of pandemic interventions, particularly in terms of mortality. The inclusion of the "Deceased" compartment is essential for capturing the full scope of the disease's impact, which is central to our study’s objectives. The use of SIRD-type models to study epidemics has been very popular for decades~\cite{daley2001epidemic, brauer2019mathematical, fernandez2022estimating, calafiore2020time, Chatterjee}, given by coupled differential equations, and does not assume any spatial structure for the population. However, a more convenient and powerful multi-agent simulation (MAS) approach can be applied to various cases, assuming a spatial structure of ﬁnite population size~\cite{hirose2013pandemic, liu2010integration, fujita2022determination, uhrmacher2018multi, siebers2008introduction, tanimoto2019evolutionary, tanimoto2021sociophysics}. Our numerical simulations in a lattice (representing 2D regular graphs) explore deeply into the intricate web of infection propagation, recovery trajectories, and mortality rates, unfolding a narrative that suggests that, while quarantine undoubtedly delays the peak of infections, it does not constitute an impenetrable barrier to halt the relentless spread of infectious diseases.

Studies found that targeted vaccination can significantly reduce the pandemic peaks~\cite{scoglio2010efficient, nunner2022prioritizing}. Our multi-agent simulation (MAS) approach also shows that vaccination can emerge as an effective countermeasure in order to halt the rapid progression of the pandemics~\cite{frederiksen2020long, graeden2015modeling, sanz2022modelling, yang2019efficient}. The findings show how higher levels of immunity in a population can provide a stronger defense against uncontrolled disease transmission~\cite{schneider2011suppressing, torjesen2021covid}. We investigate the potential of immunity to not only reduce infection peaks but also to navigate the transition from severe outbreaks to a managed endemic state using numerical simulations. 

Few works consider local isolation of infectious agents, which is expected to prevent the most infections~\cite{herrera2019local, paltra2024local, coman2023social}. However, there is a lack of a statistical-based approach to epidemiological modeling that considers local or targeted isolation of infectious agents. We demonstrated in our MAS model that local isolation of infectious agents through contact inhibition with a probability below the percolation threshold presents an effective targeted control strategy~\cite{sander2003epidemics, croccolo2020spreading, miller2009percolation}. This approach offers potential resource efficiency when compared to broad-spectrum lockdowns or large-scale vaccination campaigns. Furthermore, it can be practically implemented in a localized manner, making it well-suited for managing specific outbreaks or regions where the infection is concentrated.

The organization of the paper is as follows. In Section~\ref{ab_model}, the agent-based model is defined. Section~\ref{results} presents and analyzes the results of these simulations. Initially, we illustrate the progression of the infection without any restrictions or countermeasures. Subsequently, we analyze this scenario by implementing movement restrictions via a lockdown. Next, we examine the impact of vaccination on the epidemic when a lockdown is not in place. Finally, we demonstrate how the local isolation model via bond dilution emerges as an effective countermeasure to the disease. Section~\ref{summary} summarizes our findings. 

\section{Agent-Based Model}\label{ab_model}
In this study, we employed an individual-based Multi-Agent Simulation (MAS) model~\cite{tanimoto2019evolutionary,tanimoto2021sociophysics} to simulate the spread of an infectious disease within a population. In the MAS model, the underlying network represents physical contact between individuals, which is crucial for simulating disease transmission. Each node in the network corresponds to an individual, and links between nodes represent potential physical interactions. These interactions are typically with neighbors or individuals that people come into contact with in their daily lives, such as family members, coworkers, or neighbors. The model assumes a time-constant network, where the structure remains consistent over the simulation period. The MAS approach was chosen because it provides a detailed and flexible framework for simulating individual behaviors and interactions. MAS models allow for the inclusion of diverse factors such as spatial distribution, movement, and localized interactions, which are challenging to model using traditional ODEs.

\begin{figure}[!htbp]
    \centering
    \includegraphics[width=0.8\columnwidth]{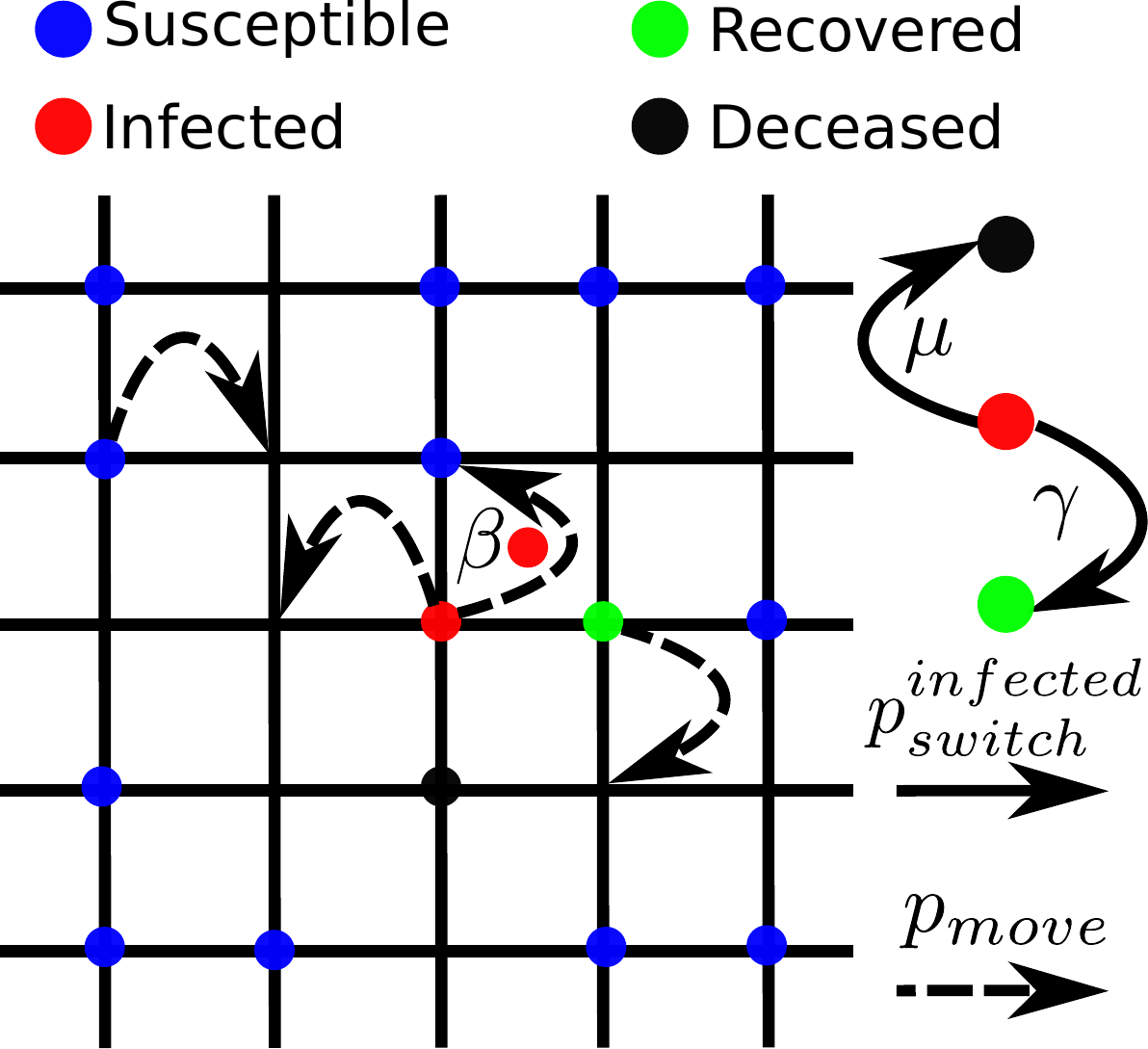}
    \caption{(Color online) In this illustration, we depict the agent-based simulation model on a square lattice. Lattice nodes can be either vacant or occupied by susceptible individuals ($S$), infected individuals ($I$), recovered individuals ($R$), or deceased individuals ($D$), shown as blue, red, green, and black circles, respectively. Susceptible individuals can move to neighboring vacant nodes and can become infected if an infected individual occupies their neighboring nodes. Infected individuals can switch to either recovered or deceased states and also can infect susceptible individuals at neighbor nodes. Recovered individuals can move freely without interaction.}
    \label{fig:fig_shematic}
\end{figure}

Let us consider $N$ agents moving on a 2D square lattice with linear dimension $L$ subject to periodic boundary conditions. Each lattice site represents a node, and for our model, $N\le L^2$ is always satisfied. A schematic representation of the model is shown in Fig.~\ref{fig:fig_shematic}. such as local isolation or targeted quarantines, can be effective in such contexts. Let $\mathscr{S}_{i,j}(t)$ denote the state of the node at position $(i,j)$ at time $t$. 
\begin{align*}
\mathscr{S}_{i,j}(t) = 0 & : \text{Vacant node.} \\
\mathscr{S}_{i,j}(t) = 1 & : \text{Node occupied by a susceptible individual.}\\
\mathscr{S}_{i,j}(t) = 2 & : \text{Node occupied by a infectious individual.} \\
\mathscr{S}_{i,j}(t) = 3 & : \text{Node occupied by a recovered individual.}\\
\end{align*}
The dynamics of the agent-based model represent a stochastic process studied using Monte Carlo simulation~\cite{mooney1997monte} and can be described as follows:
\begin{enumerate}
    \item Initialization: At time $t=0$, a fraction of lattice sites are randomly chosen and set to $\mathscr{S}_{i,j}(0) = 1$, representing susceptible individuals, except for one infectious agent $\mathscr{S}_{i,j}(0) = 2$ placed at the center of the lattice, which act as an epicenter of the outbreak. 
    
    \item Dynamics: We begin by constructing a probability interval ranging from 0 to 1. The sum over $\sum_{k=1}^{4}p_{\text{move}}^k$, where \( k \) representing the four lattice neighbors (right=1, up=2, left=3, and down=4), represents the probability of agent movement. Two additional probabilities, \( p_{\text{switch}}^{\text{infected}} \) and \( p_{\text{wait}} \), denote the probabilities of agent switching and no action occurring, respectively. The sum of all these probabilities satisfies the condition;
    \begin{equation}
       \sum_{k=1}^{4}p_{\text{move}}^k+p_{\text{switch}}^{\text{infected}}+p_{\text{wait}}=1 
    \end{equation}
    Subsequently, at each time step, an agent is randomly selected, and a uniformly distributed random number \( r[0-1] \) is generated. The time step is denoted as \( \delta t=1/N \), ensuring that, on average, all agents have an opportunity to move or switch within one Monte Carlo step($t$). The generated random number \( r \) is then compared with the probability interval.

    \begin{enumerate}
        \item Disease Spread: Based on the above comparison, if  $p_{\text{move}}^k$ is satisfied, the agent located at position $(i, j)$ moves to the specified direction($k$), provided that the target node is unoccupied. If the target node is occupied, a further examination of the states of both the selected agent and its agent at the target node is conducted. If a susceptible individual at position $(i,j)$ has a target node occupied by an infectious individual (with $\mathscr{S}_{i\pm1,j}(t) = 2$ or $\mathscr{S}_{i,j\pm1}(t) = 2$) or vice versa, susceptible become infected themselves with a rate $\beta$.
        \item Recovery \& Deceased: Based on the comparison, if $p_{\text{switch}}^{\text{infected}}$ is satisfied, the state of the selected agent is examined. If the agent is infected, it switches with a rate $\gamma$ to the recovered state ($\mathscr{S}_{i,j}(t) = 3$), becoming permanently immune to susceptibility or reinfection. Conversely, if the infected agent transitions to the deceased state with a rate $\mu$, it is removed from the system dynamics. If the agent's state is not infected, it remains unchanged. The distinction between Recovered and Deceased agents is integral to our model, particularly in the context of herd immunity. Recovered agents, by occupying nodes, can indeed create barriers that may alter the movement of susceptible and infectious agents. These "disease barriers" help slow down the spread of the infection by limiting the movement of susceptible and infectious agents, effectively creating pockets of immunity within the population. This is in line with the concept of herd immunity, where the presence of immune individuals (recovered agents) indirectly protects those who are not immune. In contrast, the removal of deceased agents highlights the importance of strategies aimed at reducing fatalities. This distinction allows us to simulate more accurately the dynamics of disease spread and the potential outcomes of various public health interventions.
        \item Waiting period: If $p_{\text{wait}}$ is satisfied through the comparison, the selected agent takes no action and remains in its current state.

    \end{enumerate}
    \item Updating: The lattice states $\mathscr{S}_{i,j}(t)$ evolve over time according to the rules defined above. 
    
All the relevant constant parameter values and variables are listed in Table~\ref{tableS1} and~\ref{tableS2}, respectively. The selection of parameters \(\beta\), \(\gamma\), and \(\mu\) is motivated by our previous work~\cite{Chatterjee, Chatterjee1}, which indicate that recovery and mortality events occur on a longer timescale compared to the infection progression, enabling a realistic disease progression. Hence, we configured \(\beta\) and \(\gamma\) to ensure that the disease advances sufficiently before recovery commences. The choice of \(\mu\) is set to be lower than \(\gamma\) since, in most infectious disease outbreaks, fatalities occur at a smaller scale in comparison to recoveries. On the other hand, the parameters \(p_{move}\), \(p_{switch}^{infected}\), and \(p_{wait}\) do not significantly affect the data quality. Instead, they modulate the pace of disease outbreak progression, with larger values accelerating the progression and smaller values decelerating it (see Appendix~\ref{app_disease_outbreak_varying_p}).

\end{enumerate}
\begin{table}[!htbp]
\begin{adjustwidth}{-0.0cm}{-0cm}
\centering
\begin{tabular}{ |c|c|c|c| } 
\hline
Parameters & Description & Values\\ 
\hline
$\beta$& \makecell{rate of infection} & 0.2786\\ 
\hline
$\gamma$& rate of recovery & 0.0608\\ 
\hline
$\mu$& rate of death & 0.0028\\ 
\hline
$p_{move}$& \makecell{movement probability} & 0.2\\ 
\hline
$p_{switch}^{infected}$& \makecell{switching probability} & 0.02\\ 
\hline
$p_{wait}$& \makecell{waiting probability} & 0.18\\ 
\hline
\end{tabular}
\caption{List of parameters chosen for the agent-based simulation.}
\label{tableS1}
\end{adjustwidth}
\end{table}
\begin{table}[!htbp]
\begin{adjustwidth}{-0.0cm}{-0cm}
\centering
\begin{tabular}{ |c|c|c|c| } 
\hline
Variables & Description\\ 
\hline
$L$& \makecell{Lattice linear dimension}\\ 
\hline
$N$& \makecell{Total number of agents}\\ 
\hline
$\xi$& \makecell{Lockdown probability}\\ 
\hline
$h_{if}$& \makecell{Vaccination factor}\\ 
\hline
$\lambda_{aff}$& \makecell{Total affected agents}\\ 
\hline
\end{tabular}
\caption{List of variables.}
\label{tableS2}
\end{adjustwidth}
\end{table}

To ensure the stability and robustness of our results, each numerical result presented in the manuscript is averaged over 100 independent simulation runs. This approach minimizes the effects of stochastic variations and ensures that the conclusions drawn from the data are reliable and statistically significant.

\subsection{Accounting the effects of lockdown}
In the simulation, at each time step $\delta t$, when an agent decides to move to one of the vacant neighboring nodes($k$) according to $p_{move}^k$, a random number between 0 and 1 is generated and then compared with the lockdown probability $\xi$. If the generated random number is less than or equal to $\xi$, the agent will proceed with the move. The lockdown effect is captured by adjusting the lockdown probability $\xi$ according to the desired level of restrictions in movement, where:
\begin{align*}
0 \leq \xi \leq 1
\end{align*}
A value of $\xi = 1$ reflects no movement restrictions, while $\xi = 0$ enforces a complete lockdown scenario. The population of each of the compartments $S$, $I$, $R$, and $D$ is normalized over $N$ for all the quantifications.

\section{Numerical Results of Agent-Based Simulation}\label{results}
\begin{figure}[!htbp]
    \centering
    \includegraphics[width=0.8\columnwidth]{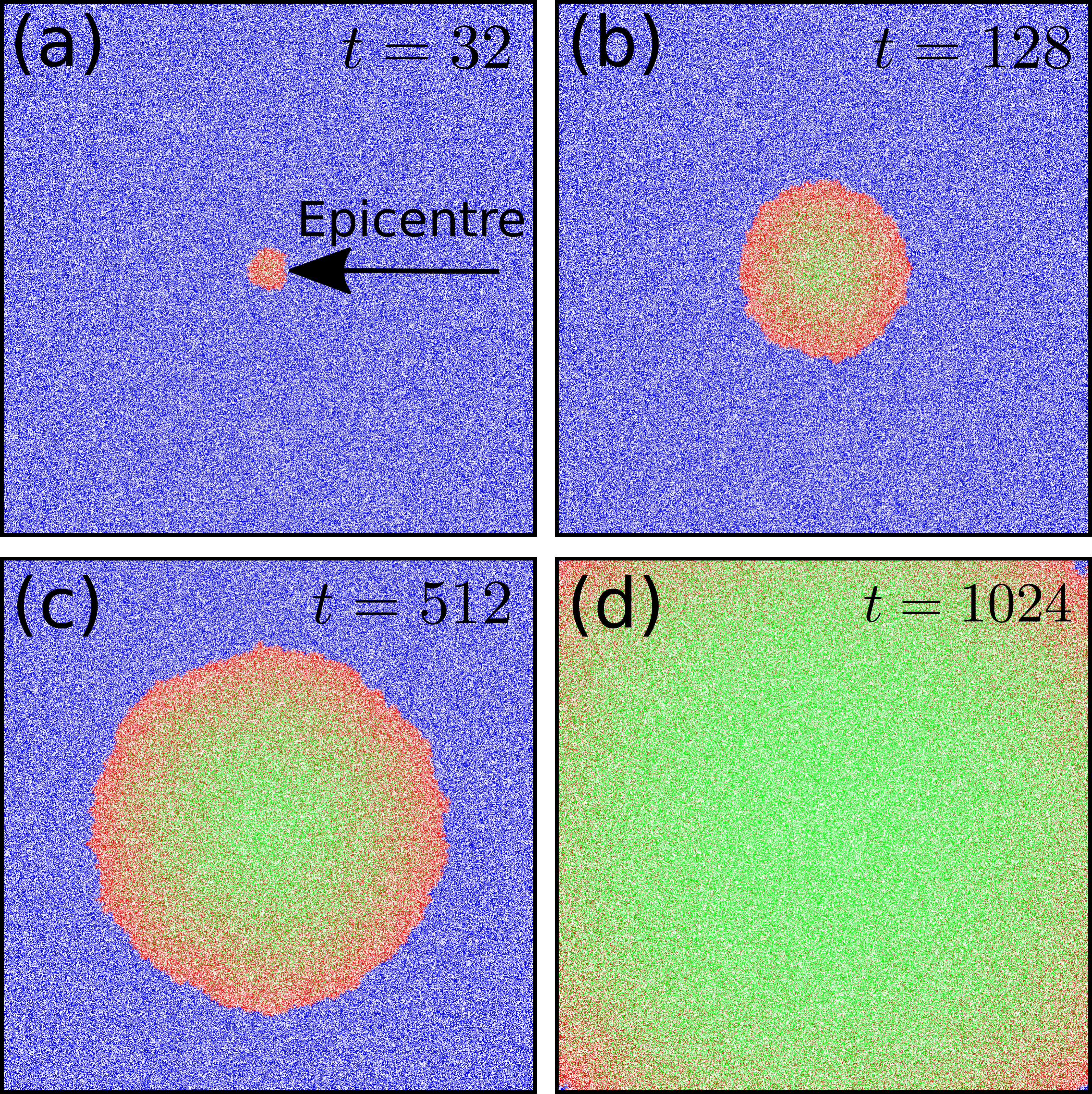}
    \caption{(Color online) Time evolution snapshots of the disease outbreak: (a) Disease starts to grow radially (red region) from the epicenter. (b-c) Infected agents recovered at the center portion (green region), and susceptible agents became infected via the infectious outer surface of the disease outbreak network. (d) Almost all the susceptible agents were infected, and most of them are recovered. [Constant parameters: $L = 1024$, $N=0.5\times L^2$, $\xi = 1$].}
    \label{fig:fig_snap}
\end{figure}

The presented series of time evolution snapshots in Fig.~\ref{fig:fig_snap} provides a compelling visual 
narrative of the temporal progression of a disease outbreak within an agent-based simulation. In the initial stages in Fig.~\ref{fig:fig_snap}(a-b), the infection emanates radially from the central epicenter, echoing the expected pattern of contagion from a point source.  As time progresses, the disease rapidly permeates the susceptible population, leading to an expanding wave of infection. As the outbreak progresses to its later stages in Fig.~\ref{fig:fig_snap}(c-d), a significant portion of the population has been affected, resulting in a notable decrease in the number of susceptible individuals with a growing number of agents transitioning to the immune or recovered state.

\begin{figure}[!htbp]
    \centering
    \includegraphics[width=1\columnwidth]{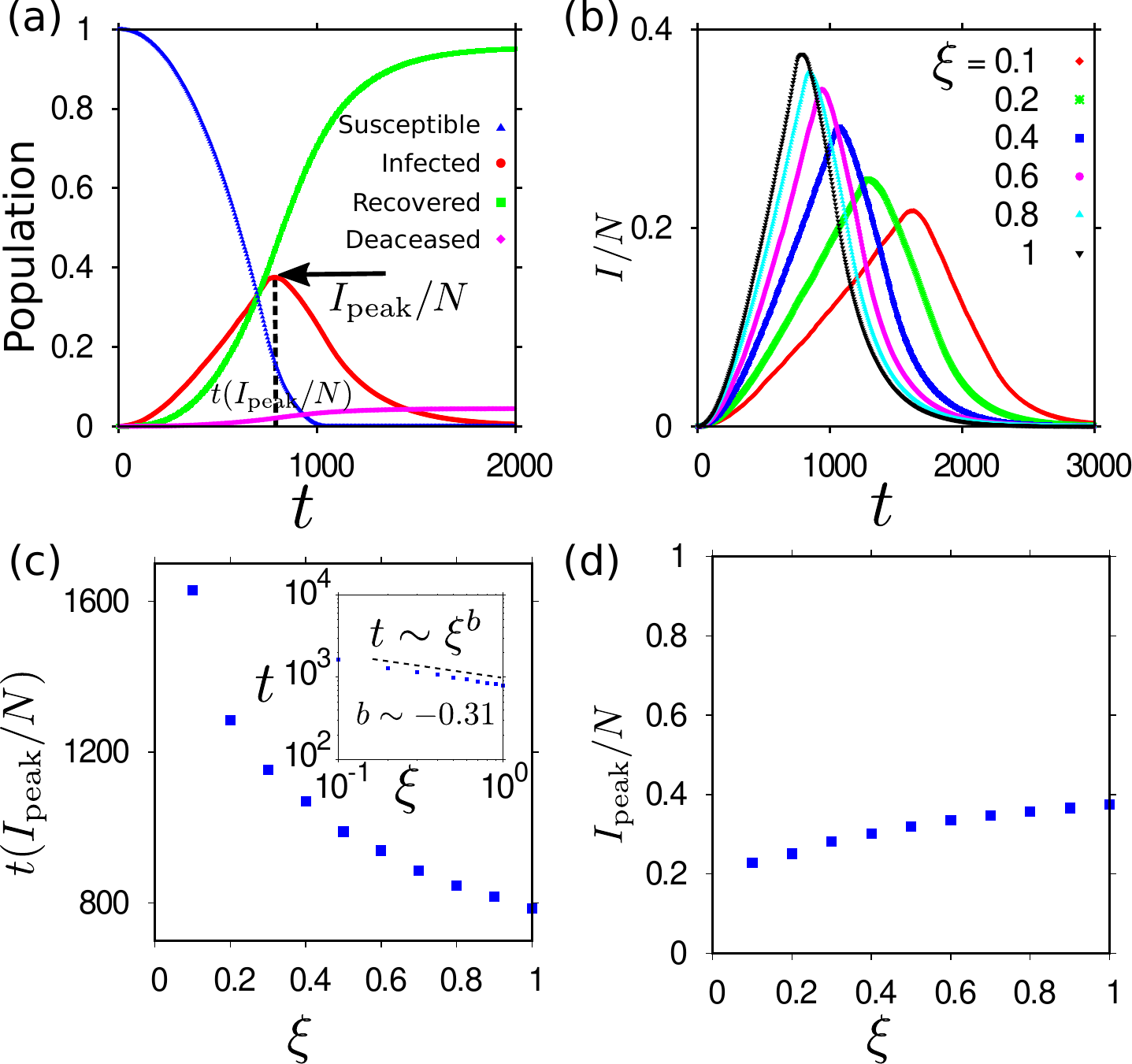}
    \caption{(Color online) (a) Trajectories of $S$, $I$, $R$, and $D$ over time for disease outbreak. (b) Variation in infectious individuals (\(I/N\)) in response to the lockdown parameter \(\xi\). (c) Effect of the lockdown parameter \(\xi\) on the time of peak infection attainment. (d) Influence of the lockdown parameter \(\xi\) on peak infection values (\(I_{\text{peak}}/N\)).}
    \label{fig:fig_quantification1}
\end{figure}

As illustrated in Fig.~\ref{fig:fig_quantification1} (a), the observed qualitative behavior mirrors that of real-world disease outbreaks~\cite{ma2020estimating} and is consistent with the results obtained from numerical solution(see Appendix~\ref{app_numerical_soln}). The susceptible individuals ($S/N$) decrease exponentially over time, indicating the rapid transmission of the disease within the population. Simultaneously, the infectious individuals ($I/N$) exhibit an exponential increase, culminating in a peak ($I_{\rm peak}/N$) that reflects the maximum prevalence of the disease. Subsequently, the infectious individuals ($I/N$) experience a decaying trend as immunity takes effect. The recovered individuals ($R/N$) demonstrate consistent growth and gradually approach a saturation value close to the total population, signifying the accumulation of immunity over time. Meanwhile, the curve representing the deceased individuals ($D/N$) accounts for the rest of the population that succumbs to the disease, indicating the unfortunate toll of the outbreak.
Varying the lockdown parameter (\(\xi\)) leads to distinguishable differences in the temporal behavior of the disease outbreak as illustrated in Fig.~\ref{fig:fig_quantification1} (b). As \(\xi\) decreases, signifying a transition from no movement restrictions to hard lockdown, distinct trends emerge in the trajectory of infectious individuals (\(I/N\)). When \(\xi = 1\) (\textit{no restrictions}), the time of peak infection ($t(I_{\rm peak}/N)$) attainment is relatively short, reflecting the rapid progression of the epidemic shown in Fig.~\ref{fig:fig_quantification1} (c). However, as \(\xi\) decreases, the time of peak infection attainment increases and follows a power law. This implies that the implementation of stringent lockdown measures can significantly delay the occurrence of the peak infection, potentially leading to a protracted outbreak. On the contrary, with decreasing \(\xi\), (\textit{hard lockdown}), the curve in Fig.~\ref{fig:fig_quantification1} (d) showcases a slight but discernible reduction in the peak infection values ($I_{\rm peak}/N$). This implies that even stringent lockdowns can only marginally mitigate the impact of the epidemic without eliminating the risk of infection.

\subsection{Impact of lockdown on deceased population \& effective reproduction number (\(R_{\text{eff}}\))}
\begin{figure*}[!htbp]
    \centering
    \includegraphics[width=2\columnwidth]{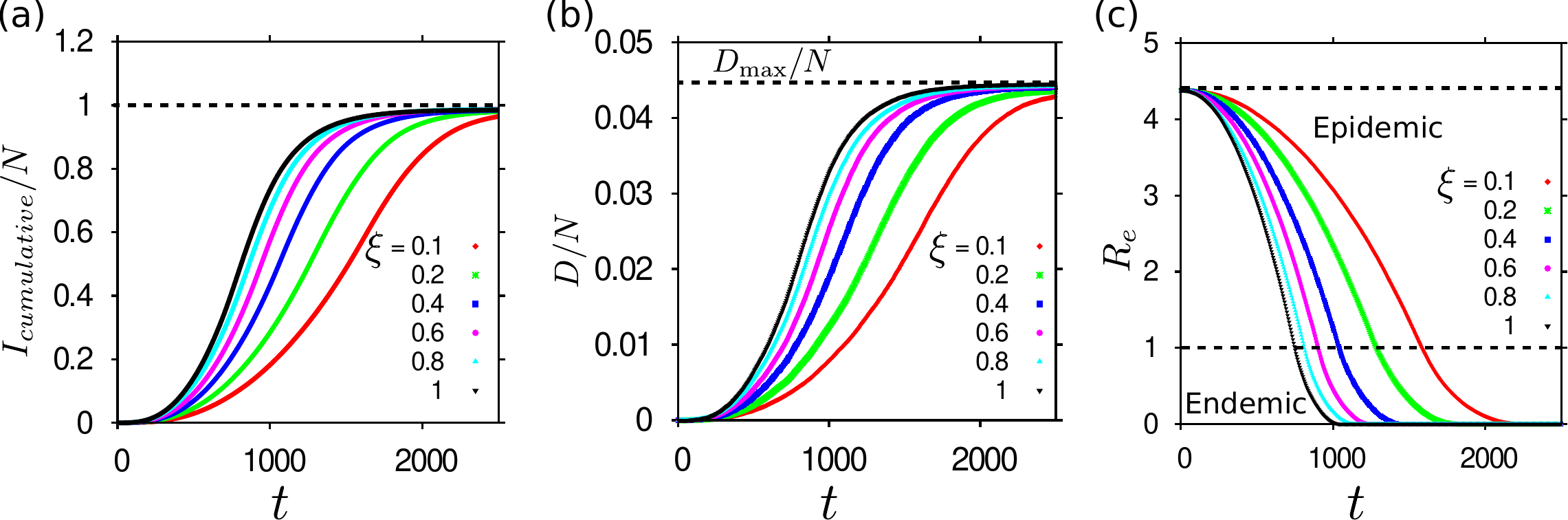}
    \caption{(Color online) (a) Cumulative number of infectious agents ($I_{cumulative}/N$) over time saturates to $1$ for different lockdown parameters \(\xi\). (b) Population of deceased individuals (\(D/N\)) as a function of $t$ for different lockdown parameters \(\xi\). (c) Effect of the lockdown parameter \(\xi\) on the effective reproduction number (\(R_{\text{eff}}\)) as a function of time.}
    \label{fig:fig_quantification2}
\end{figure*}
The cumulative number of infectious agents ($I_{cumulative}/N$), as plotted in Fig.~\ref{fig:fig_quantification2} (a), provides a clear depiction of the total infection burden over time. This plot aligns with the infection dynamics depicted in Fig.~\ref{fig:fig_quantification1} (b), where the delay in the peak corresponds to the effectiveness of lockdown in flattening the curve. However, while the peak is delayed, the overall burden (the mass under the curve) remains similar, indicating that the intervention prolongs the spread rather than reducing the total number of cases. As shown in Fig.~\ref{fig:fig_quantification2} (b), the curve demonstrates how lockdown parameters influence the temporal evolution of the deceased individuals ($D/N$). Each curve corresponds to a specific value of \(\xi\), ranging from hard lockdown (\(\xi = 0.1\)) to no restrictions (\(\xi = 1\)). All of them converge to the same saturation value ($D_{\text{max}}/N$) at different times (smaller $\xi$ takes larger time), emphasizing that while smaller \(\xi\) values can effectively halt the rapid progression of the outbreak. However, it can't significantly decrease the number of individuals who are finally deceased. 

The basic reproduction number ($R_0$) is defined as the average number of secondary infections caused by a single infected individual in a completely susceptible population~\cite{hethcote2000mathematics, delamater2019complexity, ma2020estimating}. It can be calculated using the formula:
\begin{equation}
    R_0 = \frac{\beta}{\gamma + \mu}
\end{equation}
The effective reproduction number ($R_{\text{eff}}$) accounts for changes in population immunity due to both recovery and mortality~\cite{nishiura2009effective}. It is defined as the average number of secondary infections caused by an infected individual at time $t$ in a population that is not completely susceptible. It can be calculated using the formula:
\begin{equation}
\label{R_eff}
    R_{\text{eff}}(t) = R_0 \frac{S(t)}{N}
\end{equation}
Where $S(t)$ is the number of susceptible individuals at time $t$. $R_{\text{eff}}>1$ means the epidemic could grow exponentially, leading to a larger number of infections whereas, $R_{\text{eff}}<1$ signifies disease is under control and the potential for widespread transmission is reduced. During each Monte Carlo step, each infectious agent has a single opportunity to infect a neighboring susceptible agent per time step, which constrains our ability to directly calculate \(R_{\text{eff}}\) from the simulation in terms of second-generation infections per time step. Given the constraints of the Monte Carlo simulation, we did not calculate \(R_{\text{eff}}\) directly from the number of new infections in each time step. Instead, we monitored the number of susceptible individuals present after each time step. Using this data, we applied Eq.~\ref{R_eff} to calculate \(R_{\text{eff}}\) at each time step. This approach allowed us to derive \(R_{\text{eff}}\) based on the observed changes in the susceptible population, considering the overall infection dynamics captured by the simulation.

As depicted in Fig.~\ref{fig:fig_quantification2} (c), the curve illustrates the evolution of the effective reproduction number (\(R_{\text{eff}}\)) over time for varying lockdown parameters \(\xi\), ranging from \(\xi = 0.1\) to  \(\xi = 1\). For larger values of \(\xi\), representing less stringent movement restrictions, \(R_{\text{eff}}\) falls rapidly from the initial value of \(R_0 = 4.32\), indicative of the basic reproduction number. Conversely, for smaller \(\xi\) values, \(R_{\text{eff}}\) falls comparatively in a slower rate. This scenario suggests that the epidemic is prolonged, even with stricter movement restrictions. The delayed decline of \(R_{\text{eff}}\) underscores the resilience of the epidemic, as infections continue to occur, though at a slower pace, despite efforts to limit movement and contact.

\begin{figure}[!htbp]
    \centering
    \includegraphics[width=0.8\columnwidth]{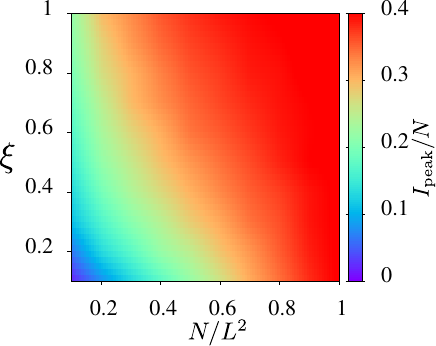}
    \caption{(Color online) Phase diagram: Fraction of total agents ($N/L^2$) vs lock-down parameter \(\xi\). The color bar indicates peak infectious individuals ($I_{peak}/N$).  }
    \label{fig:fig_phasediagram}
\end{figure}

As captured in Fig.~\ref{fig:fig_phasediagram}, the phase diagram uncovers fascinating insights into epidemic dynamics. Each data point within the diagram reflects the peak infectious individuals for specific combinations of total population fraction and lockdown parameter \(\xi\). 
Remarkably, at larger population fractions, the peak infection remains nearly consistent across different values of \(\xi\), as evidenced by the convergence of colors on the color bar. This phenomenon suggests that for more extensive community sizes, various strengths of lockdown measures have relatively comparable effects on peak infection outcomes. However, the impact of smaller \(\xi\) values comes to the forefront, notably influencing the peak of infectious individuals at lower values. This observation indicates that for smaller communities, more stringent lockdown measures (\(\xi\)) can effectively limit the peak infection, yielding a potential advantage in minimizing the impact of the outbreak.

\subsection{Incorporating vaccination factor}
For a deeper understanding of infectious disease dynamics, we introduce a new dimension – the \textit{vaccination factor} (\(h_{if}\)). It characterizes the proportion of immune individuals within the population and is quantified on a numerical scale ranging from 0 to 1, where 0 indicates the absence of immune agents and 1 represents a population entirely resilient to the disease – embodying the concept of vaccination. Within our agent-based simulation, this translates to a scenario where, at \(t=0\), a certain number ($Nh_{if}$) of the population is already immune, akin to individuals who have previously been vaccinated. These immune agents are assigned the same state as recovered individuals $\mathscr{S}_{i,j}(t) = 3$. So the rest ($N-Nh_{if}$) number of individuals act as susceptible ($\mathscr{S}_{i,j}(t) = 1$). The incorporation of \(h_{if}\) mirroring real-world circumstances where varying degrees of preexisting immunity impact disease outbreaks.

\subsection{Vaccination an important measure in infectious disease control}
\begin{figure}[!htbp]
    \centering
    \includegraphics[width=1\columnwidth]{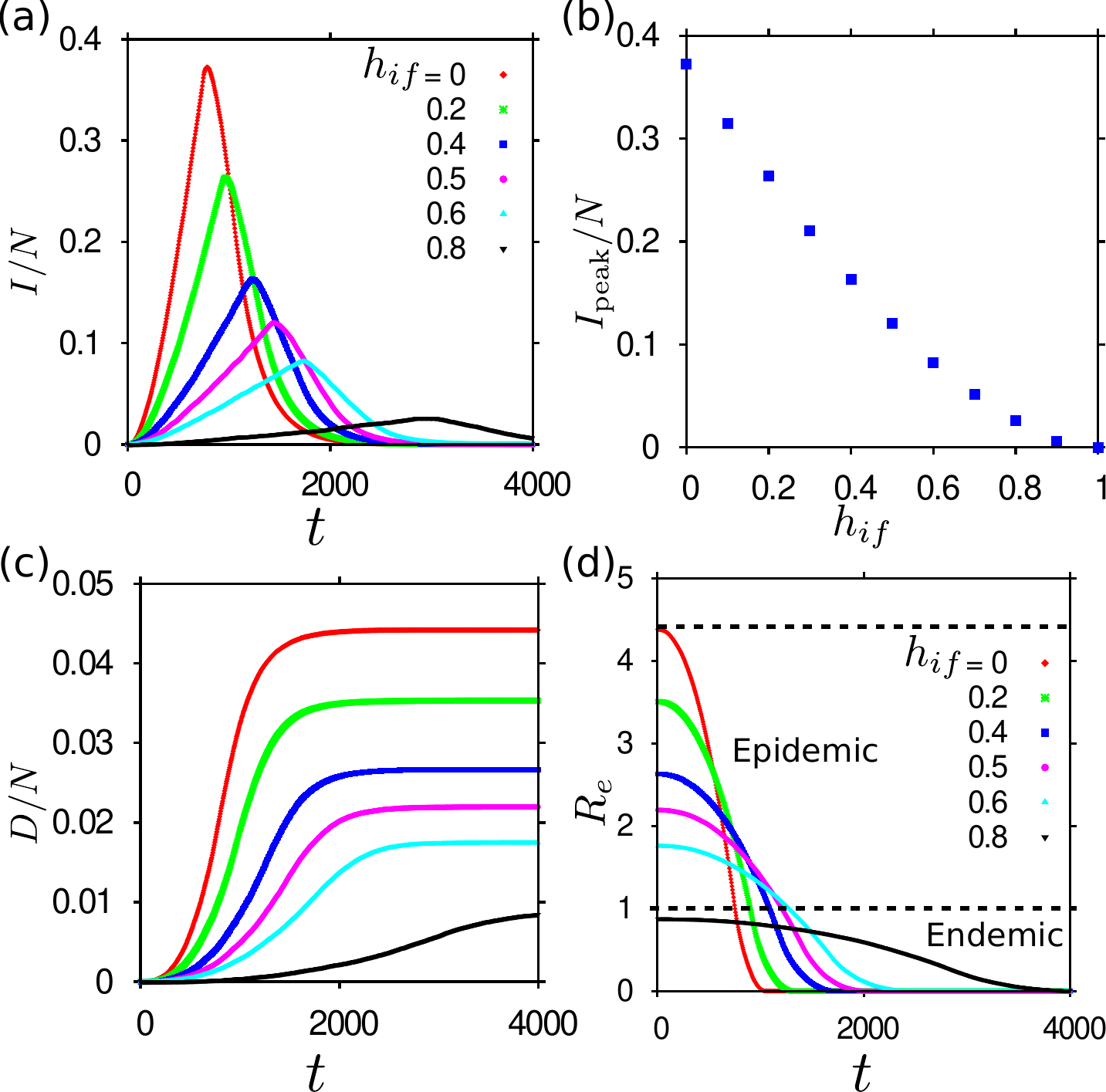}
    \caption{(Color online) (a) Impact of vaccination factor (\(h_{if}\)) on infectious individuals (\(I\)) as a function of time without restrictions (\(\xi = 1\)). (b) Peak infectious agents vs. vaccination factor (\(h_{if}\)). (c) Deceased individuals (\(D/N\)) vs. time for different \(h_{if}\) values. (d) Effective reproduction number (\(R_{\text{eff}}\)) vs. time for different \(h_{if}\) values.}
    \label{fig:fig_hifquantification}
\end{figure}

Figure~\ref{fig:fig_hifquantification} (a), portrays the variation of infectious individuals (\(I/N\)) as a function of time. Each curve represents a distinct \(h_{if}\) value, spanning the range from 0 to 0.8. With increasing \(h_{if}\) values, not only does the expected peak of infectious individuals undergo a significant delay, but it also experiences a notable reduction in magnitude in a scenario of no movement restrictions (\(\xi = 1\)).
The peak value of each curve experiences a power law descent as \(h_{if}\) is increased, as shown in Fig.~\ref{fig:fig_hifquantification} (b). This decline in power law signifies the effect of immunity on the height of the infection peak. As a larger portion of the population becomes immune, the susceptible pool contracts, resulting in a sharp reduction in the intensity of the outbreaks. Figure~\ref{fig:fig_hifquantification} (c) portraying the population of deceased individuals (\(D/N\)) over time, each curve representing different levels of the vaccination factor (\(h_{if}\)), characterized by the absence of movement restrictions (\(\xi = 1\)). All curves converge to a distinct saturation value of \(D/N\), lower for larger \(h_{if}\). This phenomenon signifies that with a greater reservoir of pre-existing immunity, the susceptible population is reduced, leading to a decrease in the overall impact on mortality. Moreover, the convergence of curves at a larger time (\(t\)) for a larger vaccination factor showcases the potential of immunity not only lowering the immediate toll of the epidemic but also prolonging the period over which the outbreak remains controlled. In Fig.~\ref{fig:fig_hifquantification} (d), where each curve represents a distinct \(h_{if}\) value, tracing the temporal evolution of \(R_{\text{eff}}\) over the course of the epidemic. Larger community immunity, reflected by higher \(h_{if}\) values, exerts a strong influence on both the basic and effective reproduction numbers. As \(h_{if}\) increases, \(R_0\) – the basic reproduction number – undergoes a noticeable reduction. Even more strikingly, \(R_{\text{eff}}\) exhibits a profound transformation. At approximately 80\% (\(h_{if}\)=0.8) immune population, a pivotal threshold is reached – the effective reproduction number dips below 1, suggesting that the disease can no longer sustain an epidemic. With time, \(R_{\text{eff}}\) descends further, eventually reaching 0.

This results in a reduced susceptible population from the outset, which indeed leads to a smaller infection curve. However, our model explicitly considers the spatial distribution of agents and the formation of "disease blockades" created by vaccinated or recovered agents. By simulating these dynamics, we demonstrate how vaccinated agents can act as barriers, influencing the movement and interaction patterns of susceptible and infected agents in a way that significantly affects the overall infection dynamics.

\subsection{Modelling local isolation in the agent-based simulation}
In the absence of both lockdown and vaccination, we introduce the concept of local isolation through bond dilution on a fixed lattice~\cite{toledano2021effects}.
\begin{figure*}[!htbp]
    \centering
    \includegraphics[width=1.9\columnwidth]{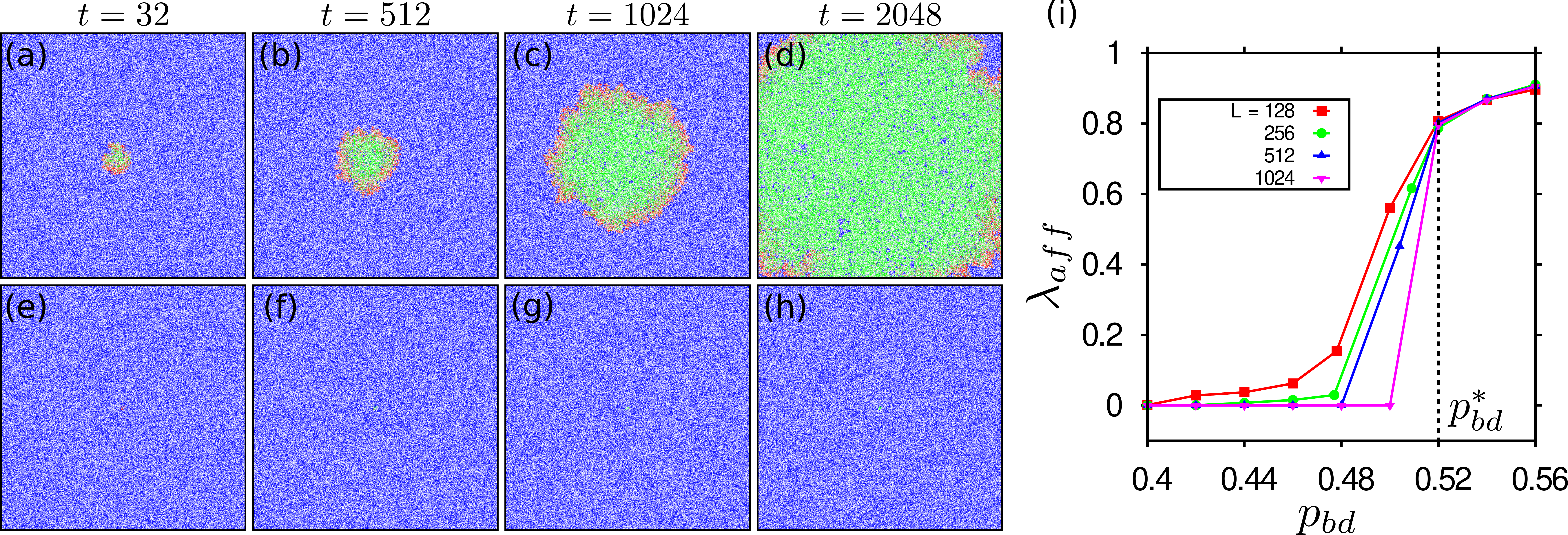}
    \caption{(Color online) Time evolution snapshots of disease outbreaks: (a-d) above the bond dilution threshold \(p_{bd}=0.53\) showing disease spread, and (e-h) just below the threshold \(p_{bd}=0.51\) where disease vanishes early, demonstrating the effectiveness of local isolation for infected agents[Fixed parameters: $L=1024$, $N=0.5\times L^2$]. (i) The percolation threshold (\(p^*_{bd} \approx 0.52 \pm 0.01\)) is obtained by calculating the number of affected agents (\(\lambda_{aff}\)) as a function of bond dilution probability (\(p_{bd}\)) for system sizes $L=128, 256, 512$, and $1024$.}
    \label{fig:fig_pbd_snap}
\end{figure*}
This model is characterized by a parameter \( p_{bd} \), which represents the probability that a bond between two neighboring nodes is present. The presence of a bond indicates a connection between the nodes, while the absence of a bond signifies isolation. The probability that a bond between nodes \( i \) and \( j \) is present is given by: 
\[
P_{ij} = \begin{cases}
    p_{bd}, & \text{if bond is present} \\
    1 - p_{bd}, & \text{if bond is absent}
\end{cases}
\]
In our model, the fundamental distinction between lockdown and local isolation lies in their targeted approach to disease containment. In the case of local isolation, only infected agents are subject to complete isolation, and they are prevented from transmitting the disease to susceptible agents. This isolation occurs if and only if the nodes connecting the infected agents to their neighbors have been diluted, rendering them ineffective for disease transmission.

\subsection{Local isolation emerges as an effective countermeasure}
We present a time evolution snapshot of the disease outbreaks for local isolation cases in Fig.~\ref{fig:fig_pbd_snap}. The top row (a-d), just above the bond dilution threshold \(p_{bd}>p^*_{bd}\), illustrates that the disease can spread effectively. However, just below the dilution threshold \(p_{bd}<p^*_{bd}\), the disease can completely vanish in the early stages, as shown in Fig.~\ref{fig:fig_pbd_snap} (e-h). This observation signifies that local isolation of infected agents can be an effective countermeasure when operating below the percolation threshold or bond dilution threshold \(p^*_{bd}\). The percolation threshold is determined by calculating the number of affected agents, \(\lambda_{aff}=\frac{R+D}{N}\), with respect to the bond dilution probability \(p_{bd}\), as depicted in Fig.~\ref{fig:fig_pbd_snap} (i). This figure reveals that as the system size increases, a sharper jump occurs at \(p^*_{bd}\), approximately at 0.52 \(\pm\) 0.01, indicating the percolation threshold.

\subsection{Fractal dimension of the disease outbreak above the percolation threshold ($p_{bd}>p^*_{bd}$)}
A fractal, in general, is a rough geometrical structure that exhibits self-similarity or scale invariance. 
\begin{figure}[!htbp]
    \centering
    \includegraphics[width=0.8\columnwidth]{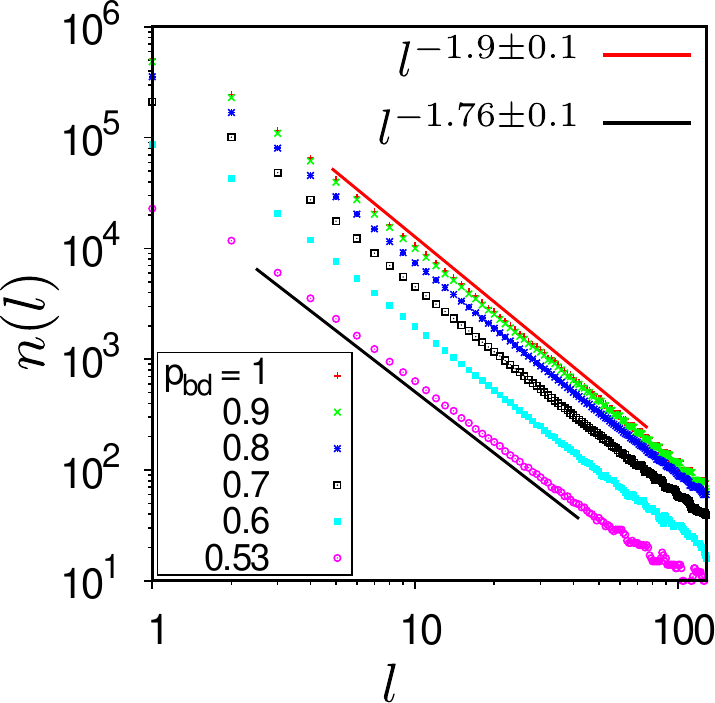}
    \caption{(Color online) Number of boxes (\(n(l)\)) as a function of box size (\(l\)) displayed on a log-log scale. Fractal dimensions (\(D_f\)) of the disease outbreak were quantified for different bond dilution levels. Above the percolation threshold, \(D_f\) exhibits slight modification, with \(D_f\) measured at \(1.9 \pm 0.1\) without dilution, decreasing to \(1.76 \pm 0.1\) at \(p_{bd} = 0.53\).}
    \label{fig:fig_fractal_SIRD}
\end{figure}
Estimating the fractal dimension provides insight into the extent of spatial influence exerted by the infection during its growth process. Therefore, it is valuable to investigate the fractal properties and quantify the structural changes in the network within an infectious environment (area covered by infectious agents [see Fig.~\ref{fig:fig_pbd_snap} (c)]). The fractal dimension (\(D_f\)) of the infectious network is determined using the box-counting method~\cite{mandelbrot1983fractal}. It is calculated as the ratio between the logarithm of the number of boxes required to cover the infectious network (\(n(l)\)) and the logarithm of the linear box size (\(l\)):
\begin{equation}
   D_f = \frac{\log(n(l))}{\log(l)} 
\end{equation}
This measurement allows us to characterize the degree of irregularity and self-similarity within the infectious network. Figure~\ref{fig:fig_fractal_SIRD} displays the number of boxes (\(n(l)\)) as a function of the box size (\(l\)) on a log-log scale. We quantified the fractal dimensions (\(D_f\)) of the disease outbreak for different bond dilution levels at a fixed $t=1024$. Above the percolation threshold, the fractal dimension (\(D_f\)) exhibits slight modification. Without dilution, the fractal dimension (\(D_f\)) is measured at \(1.9 \pm 0.1\) and decreases to \(1.76 \pm 0.1\) at \(p_{bd} = 0.53\) just above $p^*_{bd}$. This indicates that above the percolation threshold, the infection can still spread effectively even when a significant number of bonds are diluted. However, below the threshold, fractal dimension \(D_f\to 0\), signifying there is no growth of the infection (see Fig.~\ref{fig:fig_pbd_snap} (e-h)).

\subsection{Estimation of critical percolation threshold using Mean Field Theory (MFT)}
Given that 50\% of the total lattice sites are occupied and assuming an average degree \( k = 2 \) for a square lattice, we estimate the critical percolation threshold \( p_c \) using mean field theory. Below is the detailed mathematical derivation.

We consider a square lattice where 50\% of the sites are occupied. This occupation fraction reduces the effective connectivity in the lattice. In a perfect square lattice (without any vacant sites), each site typically has 4 neighbors. However, because only 50\% of the sites are occupied, the average degree is reduced. Here, we assume that the average degree \( k = 2 \) is due to the sparsity. The critical percolation threshold \( p_c \) represents the probability at which a connected cluster of infected sites first emerges and sustains itself through the lattice. In the mean field approximation, we estimate the critical threshold by considering the average connectivity of the sites:

Let the fraction of occupied sites be \( f = 0.5 \). For a site to be part of the giant connected cluster, it must be occupied and have at least one of its neighbors occupied as well. The effective degree \( k_{\text{eff}} \) of a site is the number of occupied neighbors it has. On average, each site has 4 potential neighbors, but due to the occupation fraction \( f \), only \( f \times 4 \) neighbors are likely to be occupied. With \( f = 0.5 \), we get:
\[
k_{\text{eff}} = 0.5 \times 4 = 2
\]
This justifies using \( k = 2 \) for our calculations.

The mean field theory gives us the condition that the product of the connection probability \( p \) and the effective degree \( k_{\text{eff}} \) should be at least 1 for percolation to occur~\cite{stauffer2018introduction, grimmett1999percolation, dorogovtsev2003evolution, callaway2000network}:
\[
p \times k_{\text{eff}} \geq 1
\]
Substituting \( k_{\text{eff}} = 2 \) into this equation, we obtain:
\[
p_c \times 2 = 1
\]
Solving for \( p_c \), we get:
\[
p_c = \frac{1}{2} = 0.5
\]

The mean field theory suggests that the critical percolation threshold \( p_c \) in this model, with 50\% of the lattice sites occupied and an effective degree of 2, is 0.5. This means that if the probability of connection \( p \) is greater than 0.5, a large connected cluster of infected sites is likely to emerge, leading to a sustained epidemic. The mean field theory simplifies interactions and assumes a homogeneous distribution of occupied sites and neighbors. This theoretical result is well agreement with the numerical simulations compared \( p_c \approx 0.52 \). In practice, the actual critical threshold might vary slightly due to factors like local connectivity patterns and fluctuations.

\section{Summary and discussions}\label{summary}
Our initial simulations reveal the harsh reality of uncontrolled transmission~\cite{guzzetta2020potential}, highlighting the exponential growth of infections~\cite{ma2020estimating, viboud2016generalized}. This served as a sobering reminder of the urgency for effective control measures~\cite{piret2021, VILCHES2021106564, Eugenia2013}, particularly when faced with highly contagious pathogens~\cite{mohapatra2020recent}. Introducing lockdown measures, while successful in delaying outbreaks~\cite{bogoch2015assessment, ferguson2006strategies, nussbaumer2020quarantine}, revealed- halting transmission indefinitely proved elusive. Our model underscored that lock-downs, although beneficial in buying time, cannot entirely halt the march of an epidemic~\cite{patel2020, lakhal2023, wells2021optimal}. This emphasized the need for a multifaceted approach to pandemic control. Through our simulations, we demonstrated that the presence of a larger population of pre-existing immunity via vaccination not only postpones infection peaks but also significantly diminishes their intensity. This finding reaffirms the effectiveness of immunity as a formidable tool in controlling outbreaks~\cite{frederiksen2020long, graeden2015modeling, sanz2022modelling, yang2019efficient, schneider2011suppressing, torjesen2021covid}. We also demonstrated that local isolation through bond dilution below the percolation threshold presents an effective targeted control strategy~\cite{sander2003epidemics, croccolo2020spreading}. This approach can be more resource-efficient compared to blanket lockdowns or large-scale vaccination campaigns when operating below the percolation threshold. 

In conclusion, our study illuminates the intricate interplay between disease, immunity, and intervention strategies. By investigating the details of each stage—from uncontrolled spread to lockdowns to the emergence of immunity to the localized isolation of infected agents, we provide ourselves with the knowledge necessary to confront the difficulties associated with pandemic control. Through this journey, we set a course towards a more resilient and prepared world, where science and strategy intertwine to safeguard global health~\cite{stokols1992establishing}.

Furthermore, this model can be practically implemented in a variety of real-world situations where individuals are spatially distributed in a way that limits their interactions to a local neighborhood. Examples include rural or suburban areas where residents are not densely packed, workplaces with distributed teams or even hospital settings where patients are isolated in separate rooms. The sparse lattice effectively models environments where interactions are primarily local and where movement between locations is constrained, reflecting situations where disease transmission is driven by proximity. The sparse lattice allows us to explore how localized interventions, such as local isolation or targeted quarantines, can be effective in such contexts. One could also consider the combined effect of lockdowns, local isolation, and vaccination strategies. Our primary aim was to investigate the effectiveness of localized isolation as a resource-efficient targeted control strategy. By isolating the effect of localized isolation, we can clearly identify its benefits and limitations without the confounding influences of other interventions.

\section*{Acknowledgments}
M.K. would like to acknowledge financial support in the form of a research fellowship from CSIR, the Government of India (Award No. 09/080(1106)/2019-EMR-I). M.K. acknowledges the computational facility provided by the Indian Association for the Cultivation of Science (IACS). M.K. thanks Prof. Raja Paul for the exciting discussions and insightful suggestions.

\section*{Code availability}
The simulation code was developed from scratch using the C programming language.  While the code is not publicly available under an open-source license, it can be provided upon reasonable request from the corresponding author.

\appendix
\section{Numerical Solution}\label{app_numerical_soln}
The SIRD model is a compartmental model that divides the total population($N$) into four distinct compartments~\cite{daley2001epidemic, brauer2019mathematical, fernandez2022estimating, calafiore2020time, Chatterjee}: susceptible ($S$), infectious ($I$), recovered ($R$), and deceased ($D$). The dynamics of the SIRD model can be described by the following system of ordinary differential equations:

\begin{align}\label{SIRD_equations}
\frac{dS}{dt} &= -\beta \cdot \frac{S \cdot I}{N} \\
\frac{dI}{dt} &= \beta \cdot \frac{S \cdot I}{N} - \gamma \cdot I - \mu \cdot I \\
\frac{dR}{dt} &= \gamma \cdot I \\
\frac{dD}{dt} &= \mu \cdot I
\end{align}

Where:
\begin{align*}
S(t) & : \text{Number of susceptible individuals at time } t \\
I(t) & : \text{Number of infectious individuals at time } t \\
R(t) & : \text{Number of recovered individuals at time } t \\
D(t) & : \text{Number of deceased individuals at time } t \\
N & : \text{Total population size} \\
\beta & : \text{Effective infected rate} \\
\gamma & : \text{Recovery rate} \\
\mu & : \text{Mortality rate}
\end{align*}
\begin{figure}[!htbp]
    \centering
    \includegraphics[width=\columnwidth]{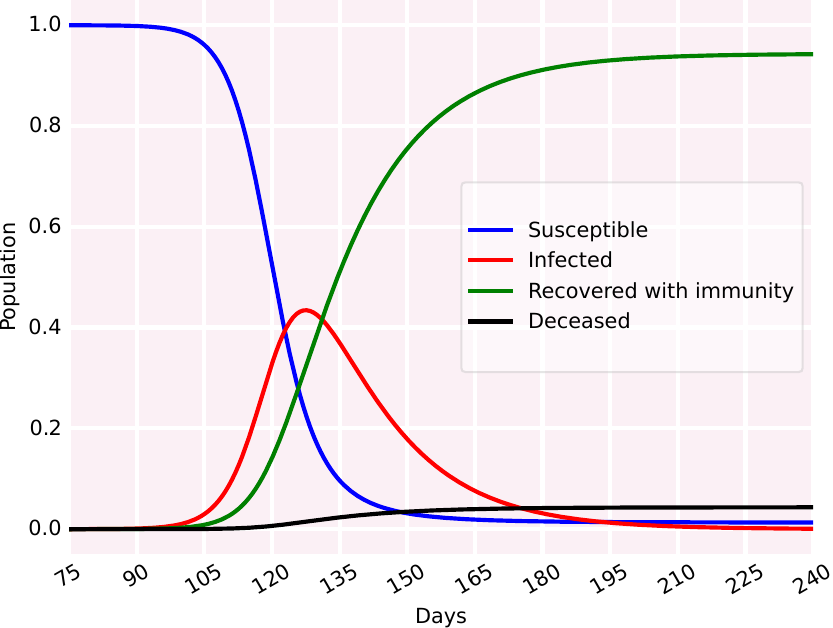}
    \caption{(Color online) Numerical solution result of disease outbreak: Trajectories of $S$, $I$, $R$, and $D$ is plotted as a function of time. }
    \label{fig:fig_numerical_solution}
\end{figure}
We numerically solve the system of differential equations~\ref{SIRD_equations}(1-4) using Euler's method~\cite{griffiths2010euler}. Let's assume initial values: $S(0) = S_0$, $I(0) = I_0$, $R(0) = R_0$, $D(0) = D_0$.  We discretize time into small intervals $\Delta t$ and update the compartments using the following equations:
\begin{align}
S_{n+1} &= S_n - \Delta t \cdot \beta \cdot \frac{S_n I_n}{N} \\
I_{n+1} &= I_n + \Delta t \cdot \left( \beta \cdot \frac{S_n I_n}{N} - \gamma I_n - \mu I_n \right) \\
R_{n+1} &= R_n + \Delta t \cdot \gamma I_n \\
D_{n+1} &= D_n + \Delta t \cdot \mu I_n
\end{align}
where the subscripts $n$ represent the time step index. 

After applying the numerical method, we plot the trajectories of $S$, $I$, $R$, and $D$ over time in Fig.~\ref{fig:fig_numerical_solution}. This provides insights into the progression of the disease outbreak.

\section{Population of infected, recovered, and deceased curve for varying $p_{switch}^{infected}$.}\label{app_disease_outbreak_varying_p}
\begin{figure*}[!htbp]
    \centering
    \includegraphics[width=2\columnwidth]{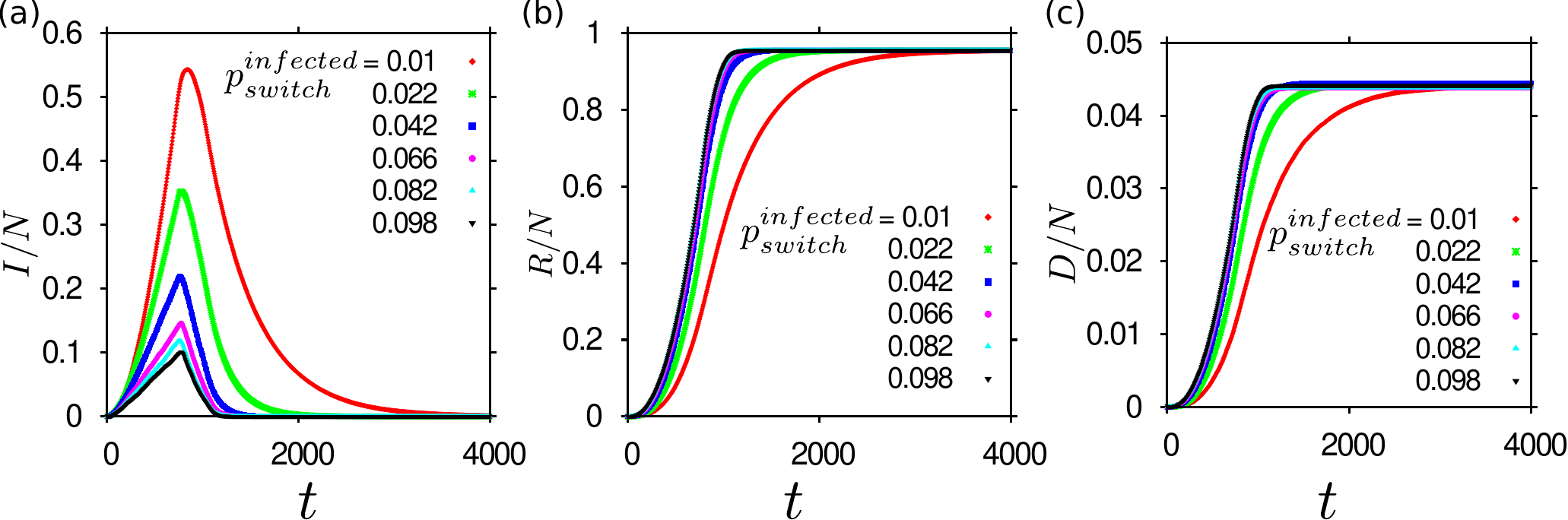}
    \caption{(Color online) Population curve of (a) Infected, (b) Recovered, and (c) Deceased agents as a function of time for varying switching probabilities \(p^{infected}_{switch}\). Fixed control parameters: [$L=1024$, $N=0.5\times L^2$].}
    \label{fig:fig_irdvst_vary_prob}
\end{figure*}
Figure~\ref{fig:fig_irdvst_vary_prob} (a-c) depict the population curves of infected, recovered, and deceased agents respectively as functions of time, each associated with varying switching probabilities \(p^{infected}_{switch}\). Qualitatively, the population of infected agents in Fig.~\ref{fig:fig_irdvst_vary_prob}(a) exhibits similar behavior across the range. However, it persists for a longer duration at smaller values of \(p^{infected}_{switch} = 0.01\). Similarly, in Fig.~\ref{fig:fig_irdvst_vary_prob}(b) and (c), the populations of recovered and deceased agents saturate later at small \(p^{infected}_{switch} = 0.01\). These findings collectively suggest that disease progression is notably slower at smaller values of \(p^{infected}_{switch}\) in comparison to larger values.

\newpage
\bibliography{biblio_Epidemic}
\end{document}